\documentstyle[12pt]{article}
\begin{document}
\addtolength{\baselineskip}{.1mm}
\input epsf
\newcommand{\ttau}{r}
\newcommand{\vev}[1]{\langle #1 \rangle}
\def\mapright#1{\!\!\!\smash{
\mathop{\longrightarrow}\limits^{#1}}\!\!\!}
\newcommand{\bigoint}{\displaystyle \oint}
\newlength{\extraspace}
\setlength{\extraspace}{2mm}
\newlength{\extraspaces}
\setlength{\extraspaces}{2.5mm}
\newcounter{dummy}
\newcommand{\be}{\begin{equation}
\addtolength{\abovedisplayskip}{\extraspaces}
\addtolength{\belowdisplayskip}{\extraspaces}
\addtolength{\abovedisplayshortskip}{\extraspace}
\addtolength{\belowdisplayshortskip}{\extraspace}}
\newcommand{\ee}{\end{equation}}
\newcommand{\figuur}[3]{
\begin{figure}[t]\begin{center}
\leavevmode\hbox{\epsfxsize=#2 \epsffile{#1.eps}}\\[3mm]
\bigskip
\parbox{15.5cm}{\small \bf Fig. 1\ 
\it #3}
\end{center} \end{figure}\hspace{-1.5mm}}
\newcommand{\fig}{{\it fig.}\ }
\newcommand{\newsection}[1]{
\vspace{15mm}
\pagebreak[3]
\addtocounter{section}{1}
\setcounter{subsection}{0}
\setcounter{footnote}{0}
\noindent
{\Large\bf \thesection. #1}
\nopagebreak
\medskip
\nopagebreak}
\newcommand{\newsubsection}[1]{
\vspace{1cm}
\pagebreak[3]
\addtocounter{subsection}{1}
\addcontentsline{toc}{subsection}{\protect
\numberline{\arabic{section}.\arabic{subsection}}{#1}}
\noindent{\bf 
\thesubsection. #1}
\nopagebreak
\vspace{2mm}
\nopagebreak}
\newcommand{\ba}{\begin{eqnarray}
\addtolength{\abovedisplayskip}{\extraspaces}
\addtolength{\belowdisplayskip}{\extraspaces}
\addtolength{\abovedisplayshortskip}{\extraspace}
\addtolength{\belowdisplayshortskip}{\extraspace}}
\newcommand{\one}{{\bf 1}}
\newcommand{\zbar}{\overline{z}}
\newcommand{\ea}{\end{eqnarray}}
\newcommand{\is}{& \!\! = \!\! &}
\newcommand{\hf}{{1\over 2}}
\newcommand{\del}{\partial}
%
%
\newcommand{\twomatrix}[4]{{\left(\begin{array}{cc}#1 & #2\\
#3 & #4 \end{array}\right)}}
\newcommand{\twomatrixd}[4]{{\left(\begin{array}{cc}
\displaystyle #1 & \displaystyle #2\\[2mm]
\displaystyle  #3  & \displaystyle #4 \end{array}\right)}}
\newcommand{\low}{{{\rm {}_{IR}}}}
\newcommand{\hi}{{{\rm {}_{UV}}}}
\newcommand{\hilo}{{{}_{{}^{\rm {}_{UV/IR}}}}}
\newcommand{\ie}{{\it i.e.\ }}
\newcommand{\gbar}{{\overline{g}}}
\newcommand{\half}{{\textstyle{1\over 2}}}
\newcommand{\tfrac}{\frac}
\newcommand{\XX}{\dot{g}}
\newcommand{\XXX}{{\mbox{\small \sc X}}}
\newcommand{\xx}{p}
\newcommand{\kappaf}{\kappa_{{}_{\! 5}}}
\newcommand{\CCC}{{\mbox{\large $\gamma$}}}
\renewcommand{\thesubsection}{\arabic{subsection}}
\renewcommand{\footnotesize}{\small}
\begin{titlepage}
\begin{center}

{\hbox to\hsize{
\hfill PUPT-1899}}

{\hbox to\hsize{
\hfill ITFA-99-40}}

\bigskip

\vspace{6\baselineskip}

{\large \sc RG-Flow, Gravity, and the Cosmological Constant}

\bigskip

\bigskip
\bigskip

{ \sc  Erik 
Verlinde\footnote{erikv@feynman.princeton.edu} 
and Herman Verlinde${}^{a}$\footnote{verlinde@feynman.princeton.edu} 
}\\[1.2cm]

{ \it Physics Department, 
Princeton University, Princeton, NJ 08544}\\[.5cm]

${}^{a}${\it Institute for Theoretical Physics, University of Amsterdam, 
1018 XE Amsterdam}\\[.5cm]

\large{

\vspace*{1.5cm}

{\bf Abstract}\\}

\end{center}
\noindent

We study the low energy effective action $S$ of gravity, induced by
integrating out gauge and matter fields, in a general class of
Randall-Sundrum type string compactification scenarios with
exponential warp factors.  Our method combines dimensional reduction
with the holographic map between between 5-d supergravity and 4-d
large $N$ field theory.  Using the classical supergravity
approximation, we derive a flow equation of the effective action $S$
that controls its behavior under scale transformations.  We find that
as a result each extremum of $S$ automatically describes a complete RG
trajectory of classical solutions.  This implies that, provided the
cosmological constant is canceled in the high energy theory, classical
flat space backgrounds naturally remain stable under the RG-flow.  The
mechanism responsible for this stability is that the non-zero vacuum
energy generated by possible phase transitions, is absorbed by a
dynamical adjustment of the contraction rate of the warp factor.

\end{titlepage}

\newpage
\newcommand{\UU}{S_E}
\newcommand{\cto}{\mu}

\newsubsection{Introduction}

In this paper we will study the low energy effective action $S$ 
of four-dimensional gravity, as obtained by integrating out all
quantum fluctuations of gauge and matter fields in a general curved 
background metric $\widehat{g}_{\mu\nu}$.  After choosing a particular
RG scale $\mu$ we can write $S$ 
as a sum 
\be 
\label{un}
\qquad \qquad S \, =
S_E + \Gamma, \qquad \qquad S_E = {1\over \kappa} \int\! 
\sqrt{-\widehat{g} }\,
\Bigl(\,2 U \, + \, \widehat{R} \Bigr) .  
\ee 
of a local Einstein action and a remaining non-local
effective action $\Gamma$, induced by the matter and gauge fluctuations
at energy scales smaller than $\mu$. 
Besides on the metric, $S$ also depends on a number of 
parameters $\phi^{I}$, representing the various masses, expectation 
values and coupling constants of this gauge and matter theory.
The seemingly inevitable presence of the potential term $U$ leads to the
well-known problem of the cosmological constant.
While one can think of various possible mechanisms or symmetries that
could ensure the absence of this term in the high
energy theory, for example (extended) supersymmetry, there is no 
real explanation known for why we do not observe the vacuum energy 
contributions produced by the various phase
transitions that take place at lower energies \cite{weinberg}.

\smallskip

In the following we will attempt to shed some new light on this
problem.  Our approach is directly motivated by the duality between
4-d large $N$ gauge theory and 5-d supergravity \cite{adscft}, as well
as by recent ideas that have appeared in the study of warped string
compactification scenarios of the type proposed in \cite{rs}
\cite{hv}.  The starting point of our discussion is the holographic
formulation of the renormalization group equations in which the RG
scale is treated as a physical extra dimension \cite{hrg1}
\cite{hrg} \cite{een}.

\smallskip

In \cite{een} it is shown that the standard RG flow of the effective
action in 4-d large $N$ field theories can be rewritten as a classical
Hamilton-Jacobi evolution equation of the 5-d supergravity action.  In
the following we will study this same idea in the context of warped
string compactifications, which can be viewed as generalizations of
the holographic duality to 4-d boundary theories with dynamical
gravity \cite{hv}.  We find that the 5-d evolution equations indeed
provide a natural extension of the 4-d Einstein equations and the
standard RG equations.  This extension seems conservative, in the
sense that most of the modifications relative to the conventional
theories seem negligible at low energies.  The most interesting
property of our equations, however, is that they appear to be
completely self-consistent for any value of the cosmological constant,
which in effect decouples from the RG-induced vacuum energy of
the matter fluctuations.  From the higher dimensional viewpoint, this
decoupling arises due to a dynamical adjustment of the contraction
rate of the warp factor, which automatically compensates for any
variations in the vacuum energy produced by the matter.  As a result
our equations are such that, assuming that $\Lambda$ is cancelled in
high energy theory, it will naturally remain zero under the RG-flow.

\smallskip

All our actual calculations will be within the 
context of classical 5-d supergravity, and thus
within the large $N$ and large coupling limit of the 4-d
field theory. We have tried, however, to formulate our results in
such a form that they represent a rather natural and conservative
extension of the standard RG framework. We believe therefore that 
a number of our conclusions will remain valid also when these
limits are relaxed. Preliminary results in this direction indeed 
indicate that there should exist natural extensions of our
formalism to other coupling regimes, as well as to finite $N$.  
Here, however, we will restrict ourselves to the simplest and
most well-controlled case.

\smallskip

Although the motivation and formulation is based on relatively recent
insights, several key elements of our proposed scenario have appeared
in earlier studies.  For example, the idea that the cosmological
constant may be cancelled by regarding our 4-d world as embedded in a
curved higher dimensional space was already suggested earlier in
\cite{rubakov}. Also the actual low energy mechanism that will be
responsible for the cancellation of the vacuum energy term is closely
related to earlier proposals for dealing with the cosmological
constant. A lucid exposition of these attempts is given in the
excellent review of Weinberg \cite{weinberg}. The idea of using the
AdS/CFT correspondence in this context was also suggested independently
by C. Schmidhuber \cite{chris}.

\smallskip

This paper is organized as follows. To set the stage, we start with 
a very brief sketch of the main idea behind our approach, and give
a short description of warped 
compactification scenarios from the point of view of the $AdS_5/CFT_4$ 
duality.
In the next few sections we then show how, along the lines of
\cite{een}, the radial evolution equations of the 5-d supergravity
action can be reformulated as an RG flow equation for the 4-d
effective action. We also describe how, as a natural by-product of 
this analysis, the 4-dimensional Einstein equations arise as a low energy 
reduction of the 5-d equations.  We then address the consequences of the 
RG-flow symmetry of the action for the cosmological constant problem, and 
show that it implies that classical flat space backgrounds, once
they are stable in the UV, also remain stable 
under the RG-flow. In the last section we draw some conclusions and
leave some open questions. Finally, in Appendix A and B we present
the form of the evolution equations for constant fields,
and compute the value of the 4-d Newton's constant for general warped 
compactifications, generalizing the RS-result \cite{rs}.

\newcommand{\delI} 
{{\partial\over \partial \phi^I}}

\smallskip

\newsubsection{Motivation: RG scale as extra dimension}

The central element in the holographic correspondence
between 5-d supergravity 
and 4-d gauge theory is the identification of the extra 5-th coordinate, 
denoted by $\ttau$, with an RG parameter of the 4-d world \cite{adscft}. 
This reinterpretation finds its origin in the following warped form of 
the 5-d metric
\be
\label{fived}
ds_{{5}}^2 = d\ttau^2\, +\, {a^2(\ttau)} \, 
\widehat{g}_{\mu\nu}\, dx^\mu dx^\nu. 
\ee 
Here $\widehat{g}_{\mu\nu}$ represents the background metric as seen by
the gauge theory.  The prefactor $a^2$ typically has an exponential
dependence on $\ttau$, growing infinitely large near the asymptotic
boundary at $\ttau \to -\infty$ and infinitesimally small at the other end.
The 4-d perspective on 5-d physics at some fixed scale thus depends on the 
radial location, in such a way that shifts in $\ttau$ have the effect of 
rescalings inside the 4-d world.

\smallskip

In the combined strong coupling and large $N$ limit of the gauge
theory, the metric (\ref{fived}) solves the classical 5-d supergravity
equations of motion, as specified via suitable boundary conditions in
the asymptotic region $\ttau \! \rightarrow \! - \infty$.  In the field
theory, these boundary conditions amount to choosing specific initial
data for the RG flow, as determined by the UV values of the various
couplings $\phi^I$. The couplings will then propagate in the
$\ttau$-direction as scalar fields of the dual supergravity.  The 5-d
geometry will thus in general deviate from its most symmetric form; it
may for example contain domain wall structures or even naked singularities 
representing specific RG flow trajectories towards non-trivial conformal
or non-conformal IR quantum field theories  \cite{hrg}.

\smallskip

If we take all $\phi^I$ independent of $x^\mu$, and $g_{\mu\nu} = a^2(\ttau)
\widehat{g}_{\mu\nu}$ with $\widehat{g}_{\mu\nu}$ some constant
curvature metric with $\widehat{R} = k$, the lagrangian of the 
5-d gravity and scalar field theory reduces to
\be
\label{lel}
L \, =\,  - {12}  a^2 \dot{a}^2\, + \, {k a^2} \, + a^4 \Bigl( 
{1\over 2} \dot{\phi}_I^2 + \, V(\phi)\, \Bigr)  .
\ee
with $V(\phi)$ some potential on the space of scalar fields.
Here we are allowed to
take $k$ independent of $\ttau$, since any possible $\ttau$-dependence
of $k$ can be absorbed in that of the scale factor $a$. Moreover,
according to the holographic postulate, this choice is necessary to
ensure that $a$ is identified with the proper RG energy-scale 
(as measured relative to the scale set by the size of the 4-d 
space-time).

\smallskip

The above lagrangian
describes the classical mechanics of a 4+1-d inflationary cosmology, 
where the radial direction, instead of the physical time direction, is
used as the time variable.  This evolution can be quite non-trivial,
and critically depends on the shape of the potential $V(\phi)$ that
drives the radial motion of the scalar fields. However, there is
always a conserved quantity due to the invariance under $\ttau$
translations. Moreover, this conserved `energy' must be set equal to 
zero by means of the Hamilton constraint of the 5-d gravity, which 
we may write somewhat suggestively as 
\be
\label{ccc}
k = -12 \dot{a}^2 + a^2\Bigl( 
{1\over 2} \dot{\phi}_I^2 - \, V(\phi)\, \Bigr).
\ee
This relation is preserved under the time-evolution derived from
(\ref{lel}), so the right-hand side also represents a conserved
quantity under the radial evolution. Since $\ttau$ represents the
holographic RG parameter, this tells us that the average curvature $k$
of the 4-d space-time is in fact an RG invariant! This somewhat
surprising fact is an immediate consequence of the holographic
dictionary, which dictates that any change in the curvature of the 4-d
slice as a function of $\ttau$ needs to be interpreted as a pure RG
rescaling with constant average curvature $k$, rather than as a
physical RG dependence of $k$.

\smallskip

The above RG-stability of $k$ indeed sounds counter-intuitive, since
normally one would expect that any RG evolution will be accompanied
with a non-zero increase in vacuum energy, which in the presence of
4-d gravity would curl up the 4-d space-time.  In the standard AdS/CFT
set-up, however, gravity is decoupled from the boundary theory. From
this perspective it is therefore not surprising that the vacuum energy
generated by the holographic RG does not directly backreact on the 4-d
geometry. However, this is not the complete story. From looking at the
explicit terms on the right-hand side of (\ref{ccc}), we can in fact
identify a specific dynamical mechanism that is responsible for this
stability, namely that any increase in the vacuum energy
produced by the RG flow of the fields $\phi^I$ automatically gets
compensated by a corresponding increase of the contraction rate of the
warp factor\footnote{A similar mechanism was also implicit in
the paper by Rubakon abd Shoposhnikov \cite{rubakov}, and independently
suggested within the present context by C. Schmidhuber \cite{chris}.}

\smallskip

In the remainder of this paper, we will explore to what extend this
idea can be used to shed new light on the cosmological constant
problem. For this, the two main questions that need to be addressed
are $(i)$ can the holographic correspondence be carried over to a
situation {\it with} dynamical 4-d gravity, and if so $(ii)$ what are
the specific modifications, relative to the standard rules
of 4-d effective field theory, that arise from this 5-d perspective 
on 4-d gravity and the renormalization group? We start with the
first question.

\smallskip

\newsubsection{Warped compactification}

In the standard AdS/CFT set-up, the boundary theory does not contain
from 4-d gravity since the AdS-space is taken to be non-compact.
Hence 5-dimensional graviton modes that extend all the way to the UV
boundary $\ttau\!  \rightarrow\!  -\infty$ are not normalizable.
However, as first pointed out by Randall and Sundrum \cite{rs}, this
situation changes as soon as one introduces a physical brane-like
structure (a ``Planck brane'') that in effect removes this infinite UV
region, thereby truncating the $\ttau$-region to a semi-infinite
range. In this case there will exist normalizable fluctuations of the
5-d metric $ds^2$ that propagate and couple as 4-d graviton modes to
the boundary field theory \cite{rs}. These graviton modes correspond
to fluctuations of the 5-d metric that preserve the warped form
(\ref{fived}), but with $\widehat{g}_{\mu\nu}$ replaced by a general
fluctuating 4-metric. The resulting strength of the 4-d gravity force
as a function of the 5-th coordinate $\ttau$ is in complete accordance
with the holographic identification of the warp factor $a$ with the RG
scale\footnote{An early suggestion that the RS-scenario \cite{rs}
might be given a holographic interpretation was made by J. Maldacena
\cite{juan}.} \cite{hv}.

\smallskip

This mechanism for including a dynamical graviton
naturally arises in a general class of string compactifications based
on type IIB orientifolds and/or F-theory \cite{hv}. 
In this case the 6-d internal 
manifold $K_6$ can carry via its topology an effective
D3-brane charge, which may be compensated by an appropriate
number of explicit D3-brane insertions \cite{warp}. 
These D3-branes wrap the 4-d
uncompactified world, and are localized as point-like objects inside
the $K_6$. Upon taking into account gravitational backreaction, this
typically leads to a warped space-time geometry of the form
\be
\label{warp}
ds^2 = a^2(\ttau) \widehat{g}_{\mu\nu}dx^\mu dx^\nu \, + 
h_{mn}(\ttau) d\ttau^m d\ttau^n.
\ee
Here $\ttau^m$ and $h_{mn}$ denote the coordinates and metric on $K_6$.

\smallskip

The most extreme type of warped compactification arises when a
relatively large number $N$ of D3-branes coalesce inside a small
sub-region inside the $K_6$ manifold \cite{hv}. In this case,
we may visualize the total ten-dimensional target space $\Sigma_{10}$ 
as obtained by gluing together two parts
\be
\label{shilo}
\Sigma_{10} = \Sigma_{\hi} \! \cup \Sigma_{\low}, 
\ee 
where $\Sigma_\low$ describes the near-horizon geometry close to 
the location of the $N$ D3-branes, and $\Sigma_\hi$ the remaining part
of the target space. In the near horizon region $\Sigma_\low$, it 
is natural to use the warp factor $a^2$ to isolate one of the 
$\ttau$ directions
to be like the extra 5-th coordinate $\ttau$ in (\ref{fived}). 
By choosing the remaining 5 coordinates $y^m$ judiciously, the metric
(\ref{warp}) can then be recast in the form 
\be
\label{warp2}
ds^2 = d\ttau^2 + 
a^2(\ttau) \widehat{g}_{\mu\nu}dx^\mu dx^\nu \, 
+  \widehat{h}_{mn}(\ttau, y) dy^m dy^n.
\ee
In this way, we have represented the total 10-d geometry as a 
$\ttau$-trajectory of 9-d geometries, where each 9-geometry is
the product of 4-d space-time with metric $\widehat{g}_{\mu\nu}$ 
and an internal 5-manifold $K_5$ described by $\widehat{h}_{mn}$. 
The near-horizon region $\Sigma_\low$ thus typically looks like 
\be
\Sigma_\low = \, M_5 \times \, K_5,
\ee
where $M_5$ is a negatively curved 5-d space with metric of the warped
form (\ref{fived}).  The 5-d space $M_5$ has a boundary $\partial
M_5 = R^4$ located at a {\it finite} radial distance $\ttau \! =\!
\ttau_0$, being the place $\Sigma_\low$ gets glued onto $\Sigma_\hi$.  
The warp-factor $a$ in this region typically traverses a large range of
values, approaching zero near the D3-branes while attaining a finite
maximum value at the boundary $\ttau \! = \!  \ttau_0$. The second
submanifold $\Sigma_\hi$ looks like the original $K_6$ with a six-ball
$B_6$ with boundary $\partial B_6 = K_5$ cut out 
\be 
\Sigma_\hi = \,
R^4 \times (K_6 \! - B_6). 
\ee 
Both submanifolds have the same boundary, equal
to $R^4 \times K_5$, so that we can indeed glue them together into one
single geodesically complete ten-dimensional manifold $\Sigma_{10}$.

\smallskip

The low energy effective field theory of this type of
compactifications is a 4-dimensional gauge theory with rank equal to
$N$, the number of D3-branes. By construction, we may identify this
field theory with the holographic dual to the IIB string theory, or
supergravity if $N$ is large enough, in the near-horizon region
$\Sigma_\low$.  An important difference with the standard AdS/CFT
set-up, however, is that at high energies the 4-d theory is augmented
to a full-fledged compactified string theory. So in particular it also
contains a dynamical 4-d graviton, that behaves exactly as the bound
state graviton in the RS world-brane scenario \cite{rs}.  Notice,
however, that in our case all fields are smooth continuous functions
of $\ttau$, so that, in contrast with the RS set-up, there is no
sharply localized brane at any finite value of the warp factor.

\newsubsection{Splitting the effective action}

We would now like to use this holographic perspective to get new
insight into the structure of the low energy effective action $S$ of
4-d gravity. To this end we will need to combine the standard techniques
of dimensional and low energy reduction with elements of the holographic
map between 5-d and 4-d physics. Besides on the 4-d metric, $S$ also
depends on various couplings $\phi^{I}$; all are allowed to vary
locally with space and time, so that $S$ encodes information about
expectation values of local gauge invariant operators ${\cal O}_{I}$
as well as of the stress-energy tensor $T_{\mu\nu}$ of the matter
theory. In addition, the metric and couplings $\phi^I$ now also
represent true dynamical degrees of freedom, whose low energy
equations of motion are prescribed by $S$.

\figuur{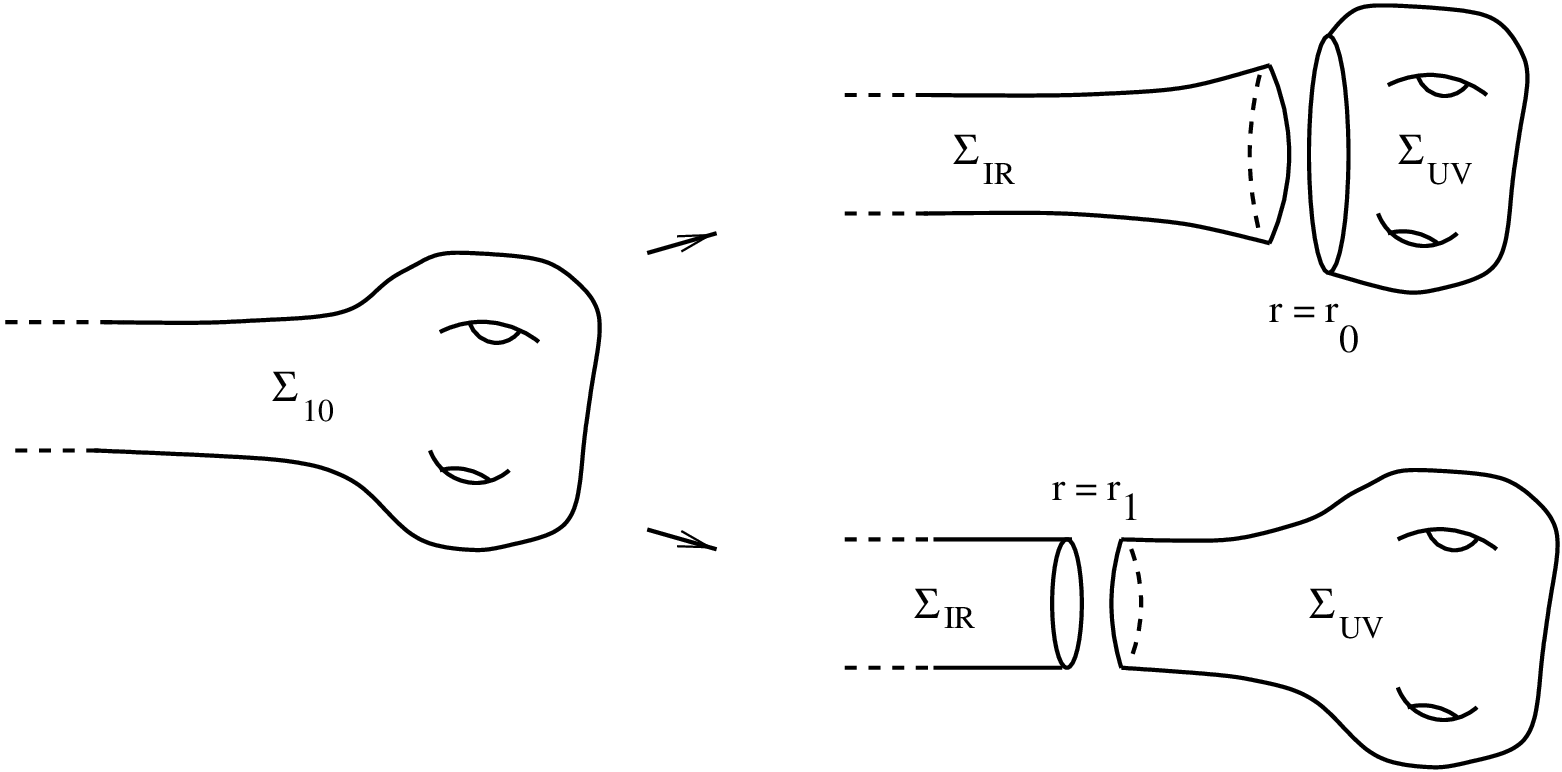}{14cm}{The total warped compactification manifold
$\Sigma_{10}$ can be split up into two submanifolds $\Sigma_\hi$ and
$\Sigma_\low$, separated by cutting along a radial location
$\ttau=\ttau_0$ close to where the near-horizon tube opens up, or at
some location $\ttau =\ttau_1$ farther inside the tube. The
corresponding split of the supergravity action $S$ has a holographic
interpretation of dividing the 4-d effective action into a high and
low energy contribution $S_\hi$ and $S_\low$, separated by a 
cut-off scale set by $\ttau_0$ or $\ttau_1$, respectively. The process
of moving the $\ttau$ location from $\ttau_0$ to $\ttau_1$ corresponds 
to performing an RG transformation in the field theory.}

\smallskip

Now imagine that we can divide up the total low energy effective action 
$S$ into a high and low energy contribution, separated by some given RG scale
$\ttau_0$, as 
\be
\label{split}
S(\phi,g) = S_\hi(\phi,g) + S_\low(\phi,g).
\ee 
Here $S_\hi$ represents the UV part of the effective action, obtained
from the original high energy action prescribed by the specific
string compactification, by integrating out all degrees of
freedom with energy larger than the IR cut-off scale set by $\ttau_0$,
and $S_\low$ is the remaining contribution to $S$ from 
all lower energy degrees of freedom.
$S_\hi$ therefore in essence describes the Einstein action of 4-dimensional
gravity, coupled to the scalar fields $\phi^I$. The values of
the couplings in $S_\hi$, such as the Newton and cosmological constant, 
are determined by their initial values as derived from the Kaluza-Klein
reduction, corrected by the effects of quantum fluctuations
down to the given cut-off scale. $S_\low$, on the other
hand, can best be thought 
of as the non-local quantum effective action of the gauge and matter system
with UV cut-off set by $\ttau_0$. 

\smallskip

Although the above split can 
be considered in a general context, it is instructive to view it as the 
holographic projection inside the boundary theory of the geometric split 
(\ref{shilo}) of the space-time manifold in the warped string compactification
scenario discussed above.
So let us for now assume that the 4-d field theory has a
sufficiently strong coupling and large enough gauge group, so that the
dual supergravity system on the warped compactification
manifold (\ref{warp}) is well approximated by its classical field
equations. In this case we can apply the standard AdS/CFT 
dictionary, and identify the 5-d scalar fields $\phi^I$
at given radial position $\ttau$ with the matter
couplings at the corresponding scale. In
particular we can identify the low-energy contribution $S_\low$ 
with the part of the classical supergravity
action coming the integral over the near-horizon region 
$\Sigma_\low$
\be
\label{slow}
S_\low(\phi, g )= \int_{\Sigma_\low} 
\!\! \! {\cal L}_{\rm sugra}.
\ee
Here the integrand is evaluated on a global classical solution,
with given boundary values $(\phi,g)$ and satisfying appropriate
asymptotic conditions near the D3-branes.  According to the standard
AdS/CFT dictionary, this non-local action $S_\low$  indeed encodes the
information of the gauge theory correlators, defined with a finite UV
cut-off. 

\smallskip

Now if we in addition assume that the boundary fields vary
sufficiently slowly along the non-compact space-time directions, and
that the $K_6$ is large enough compared to the string and Planck
scale, then we can similarly represent the high-energy action $S_\hi$
by the integral over the remaining subspace $\Sigma_\hi$
\be
\label{shi}
S_\hi(\phi, g ) = \int_{\Sigma_\hi} 
\!\! \! {\cal L}_{\rm sugra}.
\ee
Again, this integral is to be evaluated for an everywhere regular,
classical solution with boundary values $(\phi,g)$.
We can think of this action $S_\hi$ as obtained from a somewhat 
modified Kaluza-Klein dimensional reduction of
the 10-d supergravity action over the submanifold $K_6\!
-B_6$. As will become more evident later, the fact that in
contrast to the usual KK reduction, the integral runs over an internal
manifold with boundary is a reflection of the RG-dependence of the
fields on the IR cut-off of $S_\hi$.  

\smallskip

Indeed, since the coordinate location $\ttau_0$ separating 
$\Sigma_\hi$ and $\Sigma_\low$ is an adjustable parameter, 
we are free to consider the evolution of $S_\hi$ and $S_\low$ under variations 
of this location $\ttau_0$. In principle we can move $\ttau_0$ 
over the whole accessible
range of scales. The precise relation between such shifts 
and actual physical RG scale transformations is dictated by 
the shape of 4-d part of the 5-d metric (\ref{fived})
\be
\label{rescale}
g_{\mu\nu} = a^2(\ttau)\, \widehat{g}_{\mu\nu}.
\ee
The holographic interpretation of this 4-metric $g_{\mu\nu}$ is that
it measures distances in units of the RG scale, whereas $\widehat{g}_{\mu\nu}$
is defined relative to some {\it fixed} length unit such as the
fundamental string length. 
Hence, once we know the shape of the warp factor $a$ as a function of
$\ttau$, we have a well prescribed relation between $\ttau$-shifts and
4-d RG-scale transformations, since both are now directly linked with
constant Weyl-rescalings of $g_{\mu\nu}$. 
Unless explicitly stated otherwise, $g_{\mu\nu}$ will
from now on denote this RG scale dependent metric.

\smallskip

Another lesson we learn from the dual supergravity description is
that, in case the couplings $\phi^I$ vary locally with space and time,
we should anticipate that $g_{\mu\nu}$ may change under the $\ttau$
flow in other ways than just simple rescalings. General classical
trajectories in 5-d supergravity may describe parameter families of
4-d fields that, as $\ttau$ varies, can change their local shape. Via
holography, this means that the 4-d field configuration acquires a
non-trivial dependence on the RG scale at which one is looking.  Hence
rather than fixed RG-independent 4-d field configurations, we
are thus lead to consider the notion of RG-trajectories of 4-d
backgrounds.

\smallskip

Finally, it is natural to ask whether our definition of the integrals in 
(\ref{slow}) and (\ref{shi}) in fact needs to be supplemented
with a specific prescription for adding a boundary term or not.
Obviously, such a boundary term would automatically cancel in
the total sum action $S$, and should not affect the physics.
Our use of the division (\ref{split}) is to set-up a Hamilton-Jacobi 
formulation of the radial evolution equations of the 5-d
supergravity. In this context, it seems most natural to add
no boundary term at all.

\newsubsection{Radial evolution of the effective action}

Classical trajectories are selected by requiring that the 
field configurations $(\phi, g)$ on the boundary in between
$\Sigma_\hi$ and $\Sigma_\low$ must solve the equation of motion
of the total action $S$,
\be
\label{eom}
{\delta S \over \delta g^{\mu\nu}} = 0, \qquad \qquad \
{\delta S \over \delta \phi^I} = 0.
\ee 
The geometrical meaning of these equations is that they ensure
that the UV part of the RG trajectory ending at $(\phi,g)$  
joins smoothly onto the IR part of the trajectory.
To see this, it is useful to
think about $\ttau$ as a time direction, and recall the standard
result in classical field theory that for actions 
quadratic in time derivatives, there is a linear relation
between the field velocities and the variational
derivatives of the classical action evaluated at $\ttau$. 
Explicitly, in the regime where we can trust the 5-d supergravity,
the flow velocities $\dot{g}^{\mu\nu}_+$ and $\dot{\phi}^I_+$ 
coming from the UV are obtained from $S_\hi$ 
via\footnote{Throughout this paper we work in the 5-d Einstein
frame, using the 5-d Planck length as length unit.
In addition we choose the `temporal' gauge $g_{\ttau \ttau} =1$ and 
$g_{\ttau \mu} = 0$. Finally, we assume that the metric on the
space of couplings $\phi^I$ is flat, so that we can choose
it to be $\delta^{IJ}$.} 
\ba
\label{fflow}
{1\over \sqrt{-g}}\, {\delta S_\hi \over \delta g^{\mu\nu}}
\is 
{1\over 2} (\XX_{\mu\nu} - \XX^\lambda_\lambda g_{\mu\nu})_+
\\[4mm]
 {1\over \sqrt{-g}}\, {\delta  S_\hi 
\over \delta \phi^I}  \is  \, 
\dot{\phi}_{I,+}.
\ea
There exists an identical relation, but with reversed sign, 
between $S_\low$ and the flow
velocities $\dot{g}_-^{\mu\nu}$ and $\dot{\phi}^I_-$ coming from the IR. 
Using that $S = S_\hi + S_\low$, it is easy to see that 
the continuity condition on the fields and their $\ttau$-derivatives 
implies that $(\phi,g)$ must extremize the low energy action $S$.
Though seemingly self-evident, this will turn out to be a useful
observation. In contrast, for instance, it is {\it not}
necessary for classical field configurations to extremize the high
or low energy effective actions $S_\hi$ or $S_\low$.


\smallskip

It will be of importance for the following that each classical 
field configuration lies on one unique RG trajectory.  In the 
4-d field theory context, this is
because the RG flow is prescribed via a first order differential
equation, expressing $\XX_{\mu\nu}$ and $\dot{\phi}^I$ as certain
given functions of $g_{\mu\nu}$ and $\phi^I$.  Hence once we know the
initial position, the flow equation uniquely specifies the trajectory.
In the supergravity, on the other hand, this uniqueness is less
self-evident, since its equation of motion are second
order differential equations in $\ttau$. Still the same result holds. 
The reason is that we are interested in globally well-defined
classical trajectories.  Eqn 
(\ref{shi})  in principle gives $S_\hi$ as a unique functional of the 
fields, since the sub-space $\Sigma_\hi$
has no other boundaries except at the junction with $\Sigma_\low$
at $\ttau_0$.  Hence
one does not have the freedom to arbitrarily choose both the values
and velocities of the fields at its boundary. Rather, it is reasonable
to assume that the classical configuration inside $\Sigma_\hi$ is uniquely
determined by just the boundary values of the fields.

\smallskip

The 5-d classical supergravity equation of motion, that prescribes the
radial evolution of all quantities, can be most conveniently cast in
the form of a Hamilton constraint 
\be
\label{ham}
\frac {1}{4}
\Bigl(\XX^{\mu\nu} \XX_{\mu\nu}-\XX ^\mu_\mu \XX^\nu_\nu\Bigr)
+  \frac{1}{2} \dot{\phi}_I^{\, 2} 
+ {\cal L}(\phi,g) = 0
\ee
where
\be
{\cal L}(\phi,g) =
-\frac{1}{2} \, (\partial_\mu\phi_I)^{2}
+ \,  R \, + \, V(\phi) 
\ee
denotes the 4-d part of the 5-d local lagrangian density.
Upon inserting the relation (\ref{fflow}) and the similar relation
for $S_\low$, this 
constraint (\ref{ham}) takes the form of two functional identities
for $S_\hi$ and $S_\low$, which are the familiar Hamilton-Jacobi 
constraints of the canonical formalism of gravity. Introducing
the notation
\be
\label{bracket}
\, \{ \, S \, , \, S \, \}  \, \equiv \, \frac {1}{ \sqrt{-g}}
\Bigl( \, {{1\over 3}}
\Bigl(g^{\mu\nu} \frac{\delta S}{\delta g^{\mu\nu}}\Bigr)^2-\,
 \Bigl(\frac{\delta S}{\delta g^{\mu\nu}}\Bigr)^2
\, - \, \frac{1}{2}\, \Bigl({\delta S\over \delta \phi^I}\Bigr)^2 \, \Bigr)
\ee
we can write these two equations as 
\be
\label{nice}
\, \{  S_{\hi} , \, S_\hi \} \, =\, \, \sqrt{-g} \, {\cal L} \, =
\{  S_{\low} , \, S_\low \} \, .
\ee
Note that the bracket $\{\, S\, , \, S\, \}$ defines a local density,
depending on the
point $x$ at which the variational derivatives in (\ref{bracket}) are taken.
Thus the Hamilton-Jacobi equations (\ref{nice}) are local constraints.
It is further important to note that they are not conditions on the {\it value} 
of the UV and IR actions $S_\hi$ and $S_\low$ but rather functional 
identities that constrain their functional form. 

\newsubsection{RG flow in terms of beta functions}

Let us imagine that we know the high energy
action $S_\hi$ down to some given scale $\ttau$. We can then use the 
first of the Hamilton-Jacobi
equations (\ref{nice}) to integrate further down towards lower scales,
by systematically adding the infinitesimal contributions to $S_\hi$
from each integration step. This procedure is of course inspired by
the Wilsonian approach to the renormalization group.  To make this RG
interpretation more transparent, we express the UV-flow velocities as
\ba
\label{cflow}
\dot{\phi}^I_+ \is  \CCC \, \beta^I  \\[3.5mm]
\dot{g}^{\mu\nu}_+ \is  2\CCC g^{\mu\nu} \! +  \CCC\, \beta^{\mu\nu}
\label{dflow}
\ea
where $\beta_{\mu\nu}$ is defined to be traceless: $\beta^\mu{}_\mu = 0$.
Once we know the explicit form of $S_\hi$ we can use eqn (\ref{fflow})
to determine the function $\gamma$ and the `beta-functions'
$\beta_{\mu\nu}$ and $\beta^I$ as given functionals of the fields.
The factor $\gamma$ represents the flow rate of the conformal factor
of the metric, and appears as a common prefactor to ensure that the
functions $\beta_{\mu\nu}$ and $\beta^I$ describe the adjustment of
the metric and couplings induced by variations of the physical scale
defined by $g_{\mu\nu}$.

\smallskip

It is perhaps helpful to notice the obvious similarity between eqns
(\ref{cflow})-(\ref{dflow}) and the standard kinematic relations in
special relativity between the 4-velocity $\dot{x}^\mu$, as measured
with respect to the proper time, and the usual 3-velocities $\beta^i =
{d x^i/ dt}$ defined relative to the coordinate time $t\! =\! x^0$. In
this analogy, the extra coordinate $\ttau$ plays the role of the
proper time, whereas the scale factor $a$ is the analog of $t$. The
analogy goes quite a bit further, since using the definition
(\ref{cflow})-(\ref{dflow}), we can now rewrite the Hamilton-Jacobi
constraint for $S_\hi$ as the following identity for the
beta-functions
\be
\gamma^2 \Bigl( 1 - {1\over 48}
\beta_{\mu\nu}^{\,  2}
\, - \, {1\over 24} \beta_I^{\, 2} \Bigr)
\, = \, {1\over 12} {\cal L},
\label{twee}
\ee
which we recognize as the analog of the familiar relation $\gamma^2
(1-\beta^2) = {1}$ of special relativity.  From this viewpoint it is
quite tempting to speculate that for sensible RG flows, the lagrangian
${\cal L}$ on the right-hand side will always be positive, in which
case the beta-functions will have a maximum norm squared. Indeed, 
there are various indications that the standard RG-equations
of 4-d QFT follow from the 5-d evolution equations via the direct
analog of the non-relatistic limit $\beta^I <\! <1$, see also \cite{een}.

\smallskip

From combining the two equations (\ref{nice}) we can in addition
derive the following non-linear flow equation for the total low
energy effective action $S$ 
\be
\label{sflo}
\CCC \, \Bigl(\, 2 g^{\mu\nu} {\delta \ \over \delta g^{\mu\nu}} \, + \,  
{\beta}^{\mu\nu} {\delta \ \over \delta {g}^{\mu\nu}} 
\, + \, \beta^I {\delta \ \over \delta \phi^I}\, \Bigr) \, S \,
= \, \{ \, S\, ,\, S\, \}. 
\ee
The physical content of this relation is quite meaningful.
The left-hand side takes the form of a standard RG operation,
except that it contains functional derivatives, so that it
represents  a {\it local} rather than global scale variation.
From the perspective of the 4-d field theory, the right-hand side can
be thought of as an `anomalous' quantum correction; indeed, as shown in
\cite{een}, it contains among others the standard Weyl anomaly term.
Notice, however, that this right-hand side contribution vanishes for
on-shell field configurations that extremize $S$. 

\smallskip

Eqn (\ref{sflo}) tells us that $S$ is invariant under infinitesimal 
local field variations of the form
\ba
\label{delx}
\delta {\phi}_I \is \, \epsilon \, \Bigl(
\gamma\,
\beta_I +\, {1\over \sqrt{-g}} {\delta S \over \delta \phi^I}\Bigr) 
\\[4mm]
\delta g_{\mu\nu} \is \, \epsilon \,  \Bigl( \gamma \,(-2 g_{\mu\nu} + 
\beta_{\mu\nu}) + 
{1\over \sqrt{-g}}\Bigl( {\delta S \over \delta g^{\mu\nu}} - {1\over 3}
g_{\mu\nu} g^{\lambda\sigma}{\delta S \over \delta g^{\lambda\sigma}}\Bigr)
\Bigr)
\label{delg}
\ea
with $\epsilon$ some infinitesimal but otherwise arbitrarily varying
function of $x^\mu$. This invariance in effect trivializes the
dynamics of one local field degree of freedom.  When we restrict to
the space of classical solutions of $S$ the right-hand sides of
(\ref{delx}) and (\ref{delg}) reduce to the pure RG flow relations
defined in (\ref{cflow})-(\ref{dflow}), which shows that the latter
transformation act as a symmetry within the space of extrema
to $S$, {\it i.e.}  within the space of stable field
configurations. This fact will play a central role in the following.

\smallskip

One can notice a close resemblance between eqns (\ref{sflo}) and
(\ref{delx}) and the Polchinski equations for the ``exact''
renormalization group \cite{pol}, that describe the RG-scale
dependence of a Wilsonian action of the fields $(\phi, g)$ on the
cut-off length of their propagators. This similarity perhaps looks
surprising since in our case the RG flow is supposed to be generated
by integrating out the dual gauge and matter loops, instead of
gravitational degrees of freedom.  However, once we realize that
$\phi$ and $g_{\mu\nu}$ are in fact closed string modes and the gauge
and matter particles open string modes, it becomes clear that the
explanation of this correspondence should follow from the dual
equivalence between open string loops and closed string
propagators. Indeed, a Polchinski-type UV cut-off that regulates the
proper length of a closed string propagator amounts in the dual
channel to an IR cut-off on the maximal proper length of the open string
loop. It seems that this closed/open string duality lies at the heart
of the holographic interplay between gravity -- 5-dimensional {\it as
well as} 4-dimensional -- and the RG-flow induced by gauge and matter
loops. This perspective on the open string RG flow, as well as its
relation with the BV formulation of closed string field theory, will
be worked out in more detail in a future publication \cite{kv}.

\newsubsection{4-d Einstein equations}

In the following we will study the consequences of these 5-d evolution 
equations for the case that all fields are approximately
space-time independent.  In this regime we may truncate the action
$S_\hi$ to the leading order terms in its derivative expansion, and
keep essentially only the Einstein part of the action
\be
\label{cym}
S_\hi (\phi ,{g}) \, = \;
\int \!\! \sqrt{-{\widehat{g}}}    \, \Bigl( {{ U}} (\phi,a) \, +\, 
\Phi(\phi,a)  \widehat{R}\, - {1\over 2} \partial^\mu \phi^I M_{IJ}(\phi,a)
\partial_\mu \phi^J \Bigr) \, \equiv \, S_E(\phi, g) 
\ee
Here on the right-hand side 
we have reintroduced the physical metric $\widehat{g}_{\mu\nu}$ 
defined by extracting the overall RG scale $a$ from $g_{\mu\nu}$,
as in eqn (\ref{rescale}). 
${U}$ represents the vacuum energy contribution from all
quantum fluctuations with energies above the scale set by $a$,
and $\Phi$ the inverse `Newton constant' at this scale. 
$M_{IJ}$ denotes some metric on the space of scalar fields.
The explicit $a$-dependence in (\ref{cym}) reflects 
the presence of a fundamental length scale in the problem, given by
the string scale. In the holographic parametrization this scale is
located at some given value of the conformal factor, which we may set
at $a=1$.  So if $\mu$ denotes the RG energy scale, $a$ represents the
ratio
\be
\label{mpl}
a = {\mu \over M_{s}}.  
\ee 
From now on we will replace $S_\hi$ in our notation by $S_E$,
to indicate that it represents the gravitational part of the
action.

\newcommand{\wh}{\widehat}

\smallskip

Indeed, we can use holography to identify the
remaining part of the action $S_\low$ with the quantum 
effective action of the low energy gauge and matter theory 
\be
\label{deux}
S_\low(\phi,g) = \Gamma(\phi,\widehat{g}, a),
\ee
with couplings $\phi^I$, in the background geometry defined by 
$\widehat{g}_{\mu\nu}$, and with a given IR cut-off scale set by $a$.
In other words, instead of using the 5-d supergravity description all
the way down into the IR region, we imagine that we can use the
IR 4-d field theory to specify the dependence of $\Gamma$ on
the scale, metric and couplings.\footnote{The suggestion
here is that it might be possible in certain cases to {\it
match} the field theoretic quantities (beta-functions and 
expectation values) locally with the corresponding
supergravity quantities in a region where both sides are still 
under some control. One could imagine that in this way the two
dual description can be ``patched together'' into one 
global description of a specific RG flow, in which the supergravity
provides an accurate description of the UV region and the 4-dimensional
QFT of the IR regime.}
Variations of $\Gamma$ with respect to the latter two amount to 
insertions of the stress-energy tensor and operators ${\cal O}^I$,
respectively. So in particular
\be
{1\over \sqrt{-g}} {\delta \, \Gamma \over \delta g^{\mu\nu}}
= \, \langle \, T_{\mu\nu}\,  \rangle, 
\qquad \qquad  \ \ 
{1\over \sqrt{-g}} {\delta \, \Gamma \over\delta \phi^I}
= \, \langle \, {\cal O}_I \rangle. 
\ee
From now on we will denote the low energy action $S_\low$ by $\Gamma$.

\smallskip

Via these new identifications (\ref{cym}) and (\ref{deux}),
the geometric split $S= S_\hi+S_\low$ thus amounts to
a division of the total effective action into a gravity and a 
matter contribution
\be
\label{onetwo}
S(\phi,g) \, =\,  S_E(\phi,g) \, + \, \Gamma(\phi,g).
\ee
Correspondingly, the on-shell configurations that extremize 
$S$ are identified with solutions to the Einstein equations
combined with the $\phi^I$ equations of motion
\be
{1\over 2} U\, \wh{g}_{\mu\nu} + \, \Phi\, (\wh{R}_{\mu\nu}\! - \! {1\over 2}
\wh{R} \, \wh{g}_{\mu\nu}) \, = \, \langle \, T_{\mu\nu}\, \rangle 
+T_{\mu\nu}^\phi,
\ee
\be
\mbox{\large{\raisebox{-2pt}{$\Box_{{}_{\! M}}$}}} 
\phi_I + {\partial U \over \partial \phi^I }
\, + \, {\partial \Phi \over \partial \phi^I }\,
\widehat{R} \, = \, \langle\, {\cal O}_I \, \rangle,
\ee
where $T_{\mu\nu}^\phi$ denotes the stress-energy tensor of the
$\phi$-fields.  Although these look like quite generic equations of
motion, we know there must be a flow symmetry (\ref{cflow})-(\ref{dflow}) 
that acts on its space of classical solutions. This symmetry  
can be used to show that e.g the trace of the first equation, the
equation of motion of the scale factor, can in fact be
derived by combining the other equations. In addition, using the form 
(\ref{cym}) for $S_\hi$, one can show that (to leading order in the
low energy expansion) this flow equation attains
the exact form of the Callan-Symanzik equations for (the correlation
functions derived from) the effective action $\Gamma$, including the
correction term due the Weyl anomaly. We will not review this
calculation here, but refer to \cite{een}.

\smallskip

\newsubsection{RG dependence of the vacuum energy}

The set of flow equations (\ref{cflow}) through (\ref{sflo}) can be
used to deduce the classical RG trajectories of all relevant 
quantities as a function of the scale factor of the metric.
To test the physics of these equations, we start with the simplest 
situation: in flat space and with constant fields. To achieve this, 
we temporarily take the decoupling or decompactification limit 
in which the AdS-radius is sent to infinity. In this limit the 
4-d inverse Newton constant $\Phi$ is infinite, so that we can 
consistently choose $\wh{g}_{\mu\nu}$ to be flat.

\smallskip

So the only quantities we need to keep track of are the cosmological term $U$
in $S_E$ and the low energy effective action $\Gamma$.  In addition to their
explicit dependence on the scale $a$, both quantities also acquire an extra
implicit $a$-dependence, induced by radial RG-flow.  For constant fields, the
only non-zero beta-functions are those of the couplings $\phi^I$
\be
\label{pflow}
a {d \phi^I \over da} = \beta^I.  
\ee 
To extract the total scale dependence of both $U$ and $\Gamma$, 
we again use the Hamilton-Jacobi equations 
\ba
\label{yep}
\CCC \Bigl(a {\partial \over \partial a} - \beta^I 
{\partial \over\partial \phi^I}
\Bigr) U(\phi,a) \is \, - 2 a^4 \, V(\phi) \\[4mm]
\CCC \Bigl(a {\partial \over \partial a} - \beta^I 
{\partial\over\partial \phi^I}
\Bigr) \, \Gamma (\phi,a) \is 2 \int \! \sqrt{-{\widehat{g}}}\,a^4 \,
V(\phi)
\label{yip}
\ea  
The second of these equations is valid only on-shell. 
It tells us about the response of the low energy matter theory under a small
renormalization group step.  We see that besides the RG adjustment
(\ref{pflow}) of the couplings, there's also an additional vacuum energy
contribution, represented by the potential term on the right-hand side.  
As seen from the first eqn (\ref{yep}), however, the
same contribution that is subtracted from
$\Gamma$, is added in the same RG step to the potential term $U$.  This
suggests that $U$ accumulates the vacuum energy contribution
of all quantum fluctuations higher than the IR cut-off scale $a$.  Notice
that this cancellation between the vacuum energy contributions in the RG step
is of course nothing particularly deep, since by definition $S_E$ and
$\Gamma$ each represent complementary UV and IR parts of the total effective
action $S$.  Indeed, as shown earlier, $S$ is on-shell invariant under
(\ref{pflow}).

\smallskip

Now let us take a first look at the cosmological constant $\Lambda$
from this perspective.  $\Lambda$ is the total vacuum energy contained
in $U$ and $\Gamma$ added together.  Naively one might have expected
that it will coincide with the infra-red limit of the vacuum energy
contained in $U$, since it would seem reasonable to assume that, when
we send the RG scale $\mu$ to zero, the contribution from $\Gamma$ will
vanish.  It turns out, however, that this naive expectation is
incorrect.

\smallskip

It is clear from their respective definitions, that the total action
$S$ and the high energy part $S_E$ have rather different dynamical
roles.  Extrema of $S$ have an immediate physical significance as
classically stable field configurations, whereas $S_E$ generally has
non-zero variations that give us the RG flow velocities.  So the two
actions can coincide in the far infra-red {\it only if} all couplings
-- including the RG-dependent metric $g_{\mu\nu}$ -- 
would become static quantities
under the RG flow.  It is not difficult to convince oneself, however,
that our evolution equations in general predict that the couplings will 
continue to flow in the infra-red, and that as a result the IR
contributions to the vacuum energy will remain substantial, even
after the cut-off energy scale has been sent to zero.  We thus
conclude that there's an important numerical and physical distinction
between the cosmological constant and the total RG accumulated vacuum
energy contained in $U$.

\newsubsection{The cosmological constant}

\newcommand{\cC}{{\cal C}_{{}_{\! F}}}

Let us now return to the situation with dynamical 4-d gravity, but still
assume that all fields are $x^\mu$ independent.
We thus require that the metric $\wh{g}_{\mu\nu}$ has a constant curvature
$\widehat{R}$.  Assuming that $\widehat{R}$ is small, we may expand the total
action $S$ to leading order as (the general case with $\widehat{R}$
arbitrary is treated in Appendix A)
\be
\label{cym2}
S (\phi ,\widehat{g}, a) \, = \;
 \int \!\! \sqrt{-{\widehat{g}}}    
\, \Bigl({{ F}} (\phi,a) \, +\, {{ G}} (\phi,a)\,
\widehat{R} \Bigr) .
\ee
Here $G$ denotes the 4-d inverse Newton constant. The cosmological term
$F$ contains in addition to the  RG-dependent vacuum energy $U$ also
contributions coming from the flow velocities of the 
couplings as well as from the holographic contraction rate 
of the warp factor. As we have just shown (and see also section 2
and Appendix A), 
these extra contributions naturally adjust 
themselves to cancel the RG-variations in $U$. Indeed, as we will
see momentarily, $F$ is (on-shell) RG-invariant.  As before,
$a$ parametrizes the dependence of both $F$ and $G$ on the 
fundamental string scale.

\smallskip

The equations of motion of $\phi^I$ and $a$ read
\ba
\label{oem1}
{\partial F \over \partial \phi^I }
\, + \, {\partial G\over \partial \phi^I} 
 \, \widehat{R} \is 0, \\[4mm]
 a{\partial F \over \partial a}+ a{\partial G \over \partial a}\, \widehat{R}
\is 0.
\label{oem2} 
\ea
In addition, the flow invariance (\ref{sflo}) of $S$ reduces to 
\ba
\label{rrr}
\CCC \Bigl(a{\partial  \over \partial a} - \beta^I {\partial \over 
\partial \phi^I }\Bigr) F \is 
\{ \, F, \, F \, \}, \\[3.4mm]
\CCC \Bigl(a{\partial  \over \partial a} - \beta^I {\partial \over \partial
\phi^I}\Bigr) G \is 
2 \{ \, G, \, F \, \}
\label{sss}
\ea
where the bracket notation is defined in eqn (\ref{def}).
The consequences of this flow invariance are three-fold. 
First, it shows that not all equations in (\ref{oem1})-(\ref{oem2}) 
are independent;
e.g. once we have solved the $\phi^I$ equations of motion,
the second equation for the scale factor $a$ is automatically satisfied.
A second consequence is that, once we have found some 
extremum $(\phi_0,a_0)$ of $S$ at some given scale, we automatically 
obtain a whole 
one-dimensional trajectory of critical points traced out by the RG flow
\be
\label{trajectory}
a{d \phi^I\over da} = \beta^I  \; , 
\qquad \quad\phi^I(a_0) = \phi_0^I. 
\ee  
Thirdly, all points on this trajectory have the same value for the cosmological
constant $\widehat{R} = \Lambda$, as well as for $F$ and $G$.  
Hence in particular,
{\it if} the special solution in the UV happens to have zero
cosmological constant $\Lambda$, {\it then} the flow symmetry will
automatically imply the existence of stable flat space solutions that
extend all the way down into the far IR.

\smallskip

We should mention that actions with a very similar flow
symmetry have been the subject of active research in the past,
precisely because they appear to offer a promising route towards
eliminating the cosmological constant problem. These
attempts were mostly abandoned, however, because there is also an
important objection against this idea. This counter argument is most
clearly explained in section VI in Weinberg's review article
\cite{weinberg}. The main objection, as formulated in \cite{weinberg},
is based on the argument that by general covariance, the cosmological
term must necessarily be of the form 
\be
\label{simple}
\int \! \sqrt{-\widehat{g}} \, a^4 \, W(\phi)
\ee
Since a cosmological term of this form is a monotonic function in $a$, it
has no critical points whatsoever (except for the somewhat
singular value $a\! =\! 0$), unless by chance or fine-tuning $W$
happens to be zero.  Thus our argument above leading to the existence
of an RG trajectory (\ref{trajectory}) of critical points breaks down
in this case.  This is the reason why the presence of a vacuum energy
term is usually associated with an obstruction to finding a flat space
solution to the Einstein equation. 

\smallskip

There are a number of important new ingredients in our set-up, however, that
enable us to evade this ``no-go theorem''. First, it is quite clear that, 
because of the fact that $a$ represents the RG scale,
the cosmological term $F(\phi,a)$ in $S$ can have a more general dependence on
$a$ than (\ref{simple}).  This makes a crucial qualitative
difference, since it opens the possibility that non-trivial critical 
points of $F$ for which $a{\partial \over \partial a} F = 0$ can exist.  
Moreover, as we have
argued, there are good physical reasons to expect that the higher order 
dependence of $F$ on $a$ is precisely such that it allows for the
existence of a whole 
line of critical points, describing a complete RG trajectory of solutions 
to (\ref{oem1}). 

\smallskip

Another important difference, relative to previous proposals in this
direction, is that we identify this flow symmetry of $S$ with
the action of the renormalization group. Hence the fact that
the equation of motion of $S$ does not select a specific value for
the scale factor $a$ is not some unwanted instability of the
system, but rather an essential ingredient needed for this identification. 
Indeed, it would be odd if the choice of RG scale would be 
determined dynamically, rather than by hand.

\smallskip

Finally, it is important to notice that our results thus far do not
point to any prefered value for $\Lambda$. Our main result,
rather, is that we have given a new formulation of the RG
and Einstein equations which is internally consistent with any value 
for $\Lambda$, and thus {\it including} the value $\Lambda\! =\! 0$. 
Given the fundamental clash between the RG intuition and
the observational evidence of a small cosmological constant,
we consider this a useful step forward.

\newsubsection{Flat space stability}

\smallskip

The central open question therefore, is whether it is possible to
choose {\it natural} initial conditions in the far UV for which
$\Lambda=0$. Rather than attempting to give a conclusive answer this
question, let us make some general remarks.  Imagine we start with
some stable supersymmetric compactification of 10-dimensional string
theory. Eventually, this theory should want to break supersymmetry at
lower scales. One could imagine two ways in which this may happen:
either the planckian theory already contains soft non-supersymmetric
terms, or there could be some dynamical mechanism that breaks
supersymmetry via non-linear dynamics at some lower scale. In the
first type of scenario, it seems quite clear that the planckian theory
in effect already contains a non-zero cosmological term. The
RG-stability of $\Lambda$ will not help much in this case. In the
dynamical scenario, on the other hand, it seems that this stability
may become a very effective mechanism for keeping $\Lambda$
small. Roughly, the large separation between the string and the
supersymmetry breaking scale now translates via holography into a {\it
physical} separation between the non-supersymmetric dynamics and the
``Planck brane,'' that is, the place where the AdS-type geometry
connects onto the compactification manifold $K_6 \times R^4$.  The
idea is that when this Planck brane is locally embedded in a true
supersymmetric environment, it is protected from directly ``feeling''
the vacuum energy produced away from it in the bulk region.  Hence in
this case, it seems well possible that the 5-d supergravity equations
allow solutions for which the 4-d world is (almost) flat.

\smallskip

To illustrate this point, consider as a
basic -- though rather special -- example \cite{otherwalls} the
situation where the RG flow of the low energy theory is described
by a non-supersymmetric domain wall separating two 5-d
constant curvature regions, with different bulk cosmological
constant $V_+$ and $V_-$.  We imagine that for the field theory this
means that its RG flow connects a supersymmetric ultra-violet
fixed point to a non-supersymmetric infra-red fixed point field theory. 
Hence $V_\pm = V(\phi_\pm)$
where $\phi_+$ and $\phi_-$
denote the couplings of the UV and IR fixed point, respectively.
Effective field 
theory tells us that this transition will produce a non-zero
vacuum energy. The holographic manifestation of this vacuum
energy is the intrinsic  tension of the domain wall, which 
can be deduced by considering its contribution to the 
supergravity action $S$. For a flat domain wall, extended over
a range of scales from $a_+$ to $a_-$, this becomes
\be
\Delta S \, = \, \int \! \sqrt{-\widehat{g}} \, (U_+ - U_-)
\ee
with $U_\pm = U(\phi_\pm, a_\pm)$. 
The tension of the wall is 
obtained by taking the variation with respect the scale factor
\be
\label{nono}
\int \! \sqrt{-\wh{g}} \, T \, =\,  
= \, {1\over 4}  \int \! \sqrt{-\widehat{g}} \, (a_+ 
{\partial U_+\over \partial a_+}
 - a_- {\partial U_- \over \partial a_-}).
\ee
The right-hand side is
precisely the difference between the external forces acting on
the domain wall from the right and left, respectively,
arising from its embedding in the bulk regions with curvature
$V_+$ and $V_-$. Since by assumption $\partial_I U_\pm = 0$, 
the flat space Hamilton-Jacobi equation on each side reduces to
\be
\label{yy}
{a_\pm\over 4} {\partial U_\pm\over \partial a_\pm}= 
{a^4_\pm} \, \sqrt{3{V_\pm}}.
\ee
By comparing with (\ref{nono}), this reproduces the 
special relation   
\be
\label{special}
T = a^4_+ \sqrt{3{V_\pm}} - a^4_-\sqrt{3{V_-}}
\ee
between the tension and the bulk potential 
needed for the stability of the flat domain wall \cite{rs} 
\cite{otherwalls}. Notice that
this stability does not require any fine-tuning; instead
the balance of force  condition is automatically implied by the 
5-d supergravity equations.\footnote{Our reasoning here is in fact
slightly circular, since to demonstrate the relation (\ref{special}),
we first needed to assume that a flat domain wall solution indeed exists.
The point that's being made here, however, is that it does not require any  
fine-tuning of any couplings in the potential $V$ to ensure
stability of its possible domain wall solutions.}
Evidently, the vacuum energy of the 
RG-transition in the boundary field theory, or tension in the 
domain wall, does not necessarily lead to a 4-d curvature.

\smallskip

To complete this example, we would also need to consider the region
near the ``Planck brane.'' Here one could argue that fine-tuning is
needed.  However, since this is the high energy region, we are allowed
to use supersymmetry.  There are many examples of consistent
supersymmetric warped string compactifications that produce a flat
4-dimensional space-time.  This shows that also for the Planck brane,
the 10-d supergravity equations -- combined with high energy
supersymmetry -- will automatically produce the required relation
between the external force and internal tension to ensure stability.
The point here is that the curvature of the 4-d geometry is determined
by the local behavior of the fields in the Planck region.  Hence the 
flat space stability will not be perturbed by the low energy supersymmetry
breaking, {\it provided} the non-supersymmetric domain wall solution at
lower scales is well enough localized so that it has a negligible
effect on the supersymmetric solution near the Planck brane. 

\smallskip

\newsubsection{Discussion}

\smallskip
 
In this paper we have presented a new angle on the cosmological
constant problem, based on a combination of two main ideas: $(i)$ the
holographic correspondence between 4-d QFT and 5-d supergravity, and
$(ii)$ that via warped string compactifications of the RS-type this
duality can be extended to 4-d field theories with gravity.  Our
results show that in this type of scenario, there exists a natural
dynamical adjustment mechanism that prevents the cosmological constant
from being generated along the RG flow, once it has been cancelled in
the UV.  This result appears to contradict the standard, and seemingly
well-founded, intuition about how $\Lambda$ should behave.  On the
other hand, it is clear that any new proposal for dealing with this
problem must involve a restriction on the applicability of the
standard rules of 4-d effective field theory, as well as a
sufficiently radical yet conservative modification of these rules.

\smallskip

The main difference between our set-up and the standard rule book is
that the low energy equations of motion of 4-d gravity can now be
derived only {\it after} solving the RG-flow equations induced by
integrating out the matter.  Just like in the RS-scenario, the
massless 4-d gravity fluctuations correspond to zero-mode variations
of the 5-geometry that look like $ds_5^2 = d\ttau^2 + a(\ttau)^2
\widehat{g}_{\mu\nu}(x)$.  The detailed warped shape of this RS
graviton wave function depends critically on the stress-energy
distribution, and therefore ``knows'' about the various phase
transitions that occur at all lower energy scales.  This fact is
clearly illustrated e.g. in the computation of the 4-d Newton constant
outlined in Appendix B, as well as in section 2.  Our study shows
that, by being able to adjust its form, the 4-d graviton is in effect
able to make itself insensitive to the vacuum energy distribution
spread out inside the 5-d warped geometry.

\smallskip

In spite of this new way of formulating gravitational
dynamics, our set-up still seems consistent with most
established rules of effective QFT. In particular, as outlined in section
8, we can identify a quantity $U$ that behaves just like
the vacuum energy induced by the matter fluctuations.
However, this potential $U$ {\it is not equal} to the 
cosmological term in the Einstein equations, also not in 
the far infra-red; it differs from it by terms 
arising from non-zero flow velocities of the 
couplings as well as from the contraction rate 
of the warp factor. These extra contributions naturally adjust 
themselves to cancel the RG-induced variations of the 
vacuum energy contained in $U$.

\smallskip

All our equations were derived using the 5-d supergravity
approximation.  It's indeed been one of our implicit assumptions that
the rank of the high energy gauge group and gauge coupling are large
enough to be in the right regime, at least initially. However, soon
after the non-trivial RG behavior sets in, the 5-d solution will
almost inevitably enter a strongly curved region in which the warp
factor will approach zero within a finite proper distance.  This
behavior typically produces a naked singularity, near which our
approximations certainly break down. Still we have reasons to hope
that our main result, the RG stability of $\Lambda$, will continue to
hold in this regime as well. One specific source of hope is that our
equations are suggestive of a low energy approximation of the
fundamental BV symmetry structure of closed string field theory. Via
the dual correspondence between (planar) open string diagrams and
closed string (tree) diagrams, one can indeed establish a direct
interpretation of this BV symmetry as expressing an RG invariance of
the effective action, induced by integrating out the open strings. Our
suggestion is that this closed-open string duality may provide a
fundamental explanation (that also extends to the weak coupling regime of
the gauge theory) of much of the structure that we have found here \cite{kv}.

\smallskip

The holographic interpretation of the warped geometry differs on a
number of fundamental points with that of the original RS world-brane
scenario \cite{rs}. In our approach, all physics taking place
somewhere inside the 5-d bulk region is identified with physics
happening inside our 4-d world. Conversely, 4-d phenomena that occur
at different scales in our world, are represented as spatially
separated in the extra direction. This means that there is no sharply
localized world-brane anywhere within the warped space-time.
In addition, there is no very clear distinction between Kaluza-Klein
modes and other localized excitations of the 4-d field theory.
It is therefore a subtle, but very interesting, question which
possible new experimental signals one can typically associate
with scenarios of this type.

\smallskip

Finally, a relatively serious short-coming of our discussion thus
far is that we have only mentioned 5-d holography and not yet its
4-dimensional amplification, which should play an at least equally
important role. It is indeed clear that 4-d holography will put even
stronger restrictions on the validity of 4-d effective field theory 
than what we've discussed so far. Still, the incorporation of 
(at the moment somewhat better established) ideas from 5-d holography 
looks like a useful, albeit incomplete, step in the right direction.

\bigskip
\bigskip
\bigskip

\bigskip

{\noindent \sc Acknowledgements}

This work is supported by NSF-grant 98-02484, a Pionier fellowship of
NWO, and the Packard foundation. We would like to thank Christof
Schmidhuber for useful discussions and collaboration in the early
stages of this work.  We also acknowledge helpful discussions with
T. Banks, M. Berkooz, J. de Boer, S. Dimopoulos, S. Gubser, S. Kachru,
I. Klebanov, A. Polyakov, L. Randall, E. Silverstein, R. Sundrum, and
K. Skenderis.

\bigskip

\bigskip

\renewcommand{\theequation}{A.\arabic{equation}}
\setcounter{equation}{0}

\bigskip

\noindent
{\bf Appendix A: Radial dynamics for constant fields}\\[-2mm]

In this Appendix we summarize the
Hamilton-Jacobi equations for the case of constant fields.
Eqns (\ref{lel}) and (\ref{ccc}) are good starting points for analyzing
the dynamics of $a$ and $\phi^I$; here however we will instead choose
the set-up of sections 3 and 4, so that the 
analogy with the renormalization group equations remains somewhat
more transparent. So let us again introduce the actions $S_\hi= S_E$ 
and $S_\low = \Gamma$,
and a total action $S = S_E + \Gamma$. Each of these actions 
now represent functions of just the couplings $\phi^I$, the scale $a$ 
and $k$:
\be
\label{here}
S_E =  S_E(\phi, a ; k),\qquad \qquad \quad
S = S(\phi, a ; k).
\ee
The Hamilton-Jacobi constraints now reduce
to the following equations for the functions $\UU$ and $S$
\be
\label{de}
\{\, \UU, \, \UU\,  \} \, = \, a^8\, V  + \, a^6 k ,
\ee
and
\be
\label{lflo}
\gamma \Bigl(a {\partial \over \partial a} - \beta^I {\partial
\over\partial \phi^I}
\Bigr) \, S \, = \,  \{\,  S, \, S \,\}.
\ee
Here we used the notations
\ba
\label{truck}
\CCC \is {1\over 24 \, a^3} \, {\partial \UU \over \partial a}, 
\\[4mm]
\CCC \, \beta_I \is {1\over a^4 \, } \, { 
{\partial \UU \over \partial \phi^I}} 
\ea
\be
\label{def}
\{\, S, \, S\,  \} \equiv
\frac{1}{48}
\Bigl(a {\partial S \over \partial a}\Bigr)^2\! -  
{1\over 2}
\,  \Bigl({\partial S \over \partial\phi^I}\Bigr)^2.
\ee

\noindent
Note that (\ref{de}) is equivalent to 
\be
\label{fiets}
\CCC^2
\Bigl( 1- {1\over 24} \,\beta_I^2 \Bigr) \,=  \, { 1 \over 12 }\,
\Bigl(  V  + \, 
{k\over a^2} \Bigr).
\ee

\smallskip

\smallskip

The relation (\ref{de}) defines a non-linear differential 
equation from which, with given initial conditions near the Planck
region $a\! =\! a_0$, one can uniquely determine $\UU$ -- and thus 
also for $\gamma$ and beta-functions $\beta^I$ -- as functions of
$\phi$, $a$ and $k$. The detailed form of these functions 
of course critically depends on the form of the 5-d potential 
term $V$ as well as on the specific 
choice of asymptotic conditions in the UV and IR. 
Once we know these functions, however, we can then use 
(\ref{lflo}) to determine the function $S$ at all scales 
from a given initial condition. 
Then finally, after we have determined $S$,
we can distinguish the space of stable classical 
solutions $(\phi,a)$, by imposing the equations of motion
\be
\label{oem}
{\partial S\over  \partial \phi^I}(\phi,a; k) \, = \, 0 \qquad \qquad \ 
a {\partial S \over \partial a}(\phi,a; k)\, = \, 0
\ee
By virtue of eqn (\ref{lflo}), the space of solutions to these equations
of motion consists of one-dimensional trajectories, generated by 
${d \phi^I / da} = \beta^I$,
where all solutions along this flow have the same value for the
cosmological constant $k$.

\bigskip

\renewcommand{\theequation}{B.\arabic{equation}}
\setcounter{equation}{0}

\bigskip

\noindent
{\bf Appendix B: 4-d Newton constant}\\[-2mm]

In this second appendix we analyze the behavior of the Newton constant
$\Phi$ in $S_E$ under the RG-flow. 
This leads to an instructive comparison of our equations
with those of the Randall-Sundrum scenario.
We use the same notation as
in Appendix A.

\smallskip

Let us assume that we have obtained the
$\beta$ and $\gamma$ as a function of $a$. Newton's constant can then
be obtained as follows. First consider the quantity
\be
\qquad \qquad
\Phi(\phi,a) = {\partial \UU \over \partial k}(\phi,a;k = 0)
\ee
which, recalling that $k = \widehat{R}$ represents the
average curvature, can be seen to represent the coupling in front 
of the 4-d Einstein term in
$S_\hi$ at the scale set by $a$.  Using the equations (\ref{truck})
and ({\ref{fiets}) we derive that $\Phi$ satisfies the flow relation
\be
\label{step}
\CCC \; \Bigl(- a {\partial \over \partial a}  + \beta^I{\partial\over 
\partial \phi^I} \Bigr) \Phi\, = \, a^2. 
\ee
This equation determines the $a$-dependence $\Phi$ in terms of that of $\CCC$.
Eqn (\ref{step}) can then be explicitly integrated to
\be
\label{new}
\Phi(a) \, = \, \Phi(a_0) \,  
+ \, \int_{\mbox{$a$}}^{\mbox{\raisebox{2pt}{$a_0$}}} \! 
{a'\, d a'\over  \CCC(a')}.
\ee
The value of the integration constant
$\Phi(a_0)$ is determined by the specific string
compactification that describes the Planck scale physics.  

\smallskip

It is useful to compare the above result (\ref{new}) with the formula
for the 4-d Newton constant in the Randall-Sundrum compactification 
scenario \cite{rs}. In the set-up of \cite{rs}, the 4-d Newton constant is
expressed via a similar integral as (\ref{new}) with $\Phi(a_0)= 0$
and with $\CCC(a) = $ const. The location $a_0$ in their case corresponds
to the location of the `Planck brane' that cuts off the asymptotic
AdS-region. The above expression, however, also applies
to more general ways of cutting off the AdS-space by means
of the warped string compactifications described in section 2.
In this case, $a_0$ could be chosen to represent the place where the
AdS-space is glued into the compactification geometry $R^4 \times
K^6$. The integration constant $\Phi(a_0)$ will then be a finite
number, determined by the 10-d Newton constant $\kappa_{{}_{\! 10}}$
and the volume of the $K_6$ with the six-ball $B_6$ removed, via
\be
\qquad \ \ 
\Phi(a_0) = {V'_6 \over \kappa_{{}_{\!10}}}
\qquad \qquad \ 
V'_6 =  {\mbox{\rm Vol}}(K_6\! -\! B_6) .
\ee
This relation directly follows from the expression (\ref{shi}) for $S_\hi$.
Notice that the total $a_0$-dependence exactly cancels in the formula 
for $\Phi(a)$, since small shifts in the endpoint of the $a$-integration
are compensated by small shifts in radius of the six-ball $B_6$ inside $K_6$. 
This cancellation in ensured by the relation between the 5-d and 10-d
Newton constant ${1/\kappaf} = {V_5/\kappa_{{}_{\! 10}}}$ with 
$V_5 = \mbox{\rm Vol}(K_5)$.

\smallskip

The true 4-d Newton constant $\kappa$, as seen at long distances, is the 
IR fixed point value of $1/\Phi$. Hence
\be
\label{kkp}
{1\over \kappa} = \,  {V'_6 \over \kappa_{{}_{\!10}}} \, 
+ \,   
\,  \int_{0}^{\mbox{\raisebox{2pt}{$a_0$}}} 
 {a\, d a\over   \CCC(a)}.
\ee
This is a finite result, because $\CCC(a)$ will always remain non-zero,
and usually will even grow very large, in the IR region $a\to 0$.
It thus seems very plausible that (\ref{kkp}) indeed expresses the
physical Newton constant of the low energy action $S$.

\renewcommand{\Large}{\large}

\bigskip

\noindent

\begin{thebibliography}{99}


\bibitem{weinberg}
S. Weinberg, {\it The cosmological constant problem}, Rev.Mod.Phys.61
(1989), 1-23.
\bibitem{adscft} J. Maldacena, 
{\it  The large N limit of supercomformal field theories and supergravity},  
Adv. Theor. Math. Phys. 2 (1998) 231,
hep-th/9711200;\\
S. Gubser, I. Klebanov and A. Polyakov, {\it Gauge Theory Correlators
from Non-Critical String Theory}, Phys.Lett.B428 (1998) 105, hep-th/9802109;\\
E. Witten, {\it Anti-de Sitter Space and Holography},
Adv.Theor.Math.Phys.2 (1998) 253, hep-th/9802150;\\
L. Susskind and E. Witten, {\it The Holographic Bound in Anti-de Sitter Space},
hep-th/9805114.
\bibitem{rs}L. Randall and R. Sundrum, {\it A Large Mass Hierarchy from a
Small Extra Dimension}, hep-ph/9905221;
{\it  An Alternative to Compactification}, hep-th/9906064;\\
M. Gogberashvili,  {\it
Hierarchy problem in the shell-Universe model} hep-ph/9812296.
\bibitem{hv} H. Verlinde, {\it Holography and Compactification}, 
hep-th/9906182.
\bibitem{hrg1}
E.T. Akhmedov, {\it A Remark on the
 AdS/CFT Correspondence and the
 Renormalization Group  Flow},
 Phys. Lett. {\bf B442} (1998) 152,
 hep-th/9806217.\\
E. Alvarez and C. G\'omez
 {\it Geometric Holography, the Renormalization
 Group and the c-Theorem}, Nucl. Phys.
{\bf B541} (1999) 441, hep-th/9807226.\\
 V. Balasubramanian, P. Krauss, 
{\it Space-time and the Holographic Renormalization Group}, 
 Phys.Rev.Lett 83 (1999) 3605, hep-th/9903190.
\bibitem{hrg}
L. Girardello, M. Petrini, M. Porrati, A. Zaffaroni,
{\it Novel Local CFT and Exact Results on Perturbations of N=4 Super Yang
Mills from AdS Dynamics}, hep-th/9810126, and
{\it The Supergravity Dual of $N=1$ Super Yang-Mills Theory}, hep-th/9909047;\\
D. Z. Freedman, S. S. Gubser, K. Pilch, N. P. Warner,
{\it Renormalization Group Flows from Holography--Supersymmetry and a
c-Theorem}, hep-th/9906194;\\
 M. Porrati, A. Starinets, {\it RG Fixed Points in Supergravity Duals of 4-d Field Theory and Asymptotically AdS Spaces}, 
 Phys.Lett.
{\bf B454} (1999) 77, hep-th/9903241.
\bibitem{een} J. de Boer, E. Verlinde, and H. Verlinde,
{\it On the Holographic Renormalization Group}, hep-th/9912012
\bibitem{rubakov} V.A. Rubakov, M.E. Shaposhnikov, {\it Extra Space-Time
Dimensions: Towards a Solution of the Cosmological Constant} Phys.Lett. 
{\bf 125B} (1985) 372. {\it Do we live inside a Domain Wall?} Phys. Lett.
{\bf 125B} (1983) 139.
\bibitem{chris} C. Schmidhuber, private communication.
\bibitem{juan} J. Maldacena, remark at IAS seminar, May 1999.
\bibitem{warp} See in particular
K. Dasgupta, G. Rajesh and S. Sethi, {\it M Theory, Orientifolds and G-Flux}, 
hep-th/9908088  and references therein.   
\bibitem{pol} J. Polchinski, {\it Renormalization and Effective 
Lagrangians},Nucl. Phys, {\bf B231} (1984) 269. 
\bibitem{kv} J. Khouri and H. Verlinde, {\it Planar Diagrams and Closed String Field Theory},
in preparation.
\bibitem{otherwalls} K. Behrndt, M. Cvetic, 
{\it Supersymmetric Domain-Wall World from D=5 Simple Gauged Supergravity},
hep-th/9909058;\\
K. Skenderis, P. Townsend, {\it
Gravitational Stability and Renormalization-Group Flow}
 hep-th/9909070; \\
U. Ellwanger, {\it  Constraints on a Brane-World from the Vanishing of the Cosmological Constant}, hep-th/9909103;\\ 
O. DeWolfe, D.Z. Freedman, S.S. Gubser, A. Karch, 
{\it Modeling the fifth dimension with scalars and gravity}, 
hep-th/9909134.
\end{thebibliography}
\end{document}